\documentclass[11pt,oneside]{article} 

\usepackage{a4wide}
\usepackage[title]{appendix}

\usepackage{amsmath}
\usepackage{color}
\usepackage{framed}
\usepackage{tikz}
\usepackage{tikz-cd}

\RequirePackage{amsmath}
\RequirePackage{amssymb}
\RequirePackage{amsthm}
\RequirePackage{color}
\RequirePackage{url}
\RequirePackage{mdwlist}
\RequirePackage{arydshln}

\RequirePackage{rotating}

\RequirePackage[all]{xy}
\RequirePackage{graphicx}
\RequirePackage{subcaption}

\usepackage{xcolor}
\usepackage{amsmath,amsfonts,amssymb}
\usepackage{graphicx}
\usepackage{enumerate}
\usepackage{mathrsfs}
\usepackage{mathtools}
\usepackage{authblk}

\makeatletter
\newcommand{\verbatimfont}[1]{\renewcommand{\verbatim@font}{\ttfamily#1}}
\makeatother

\newcommand{\Kernel}{\mathrm{ker}}

\newcommand{\Span}{\mathrm{span}}

\newcommand{\Cokernel}{\mathrm{coker}}

\newcommand{\Cospan}{\mathrm{cospan}}

\newcommand{\Unitary}{\mathrm{U}}

\newcommand{\Complex}{\mathbb{C}}

\newcommand{\Field}{\mathbb{F}}

\newcommand{\tensor}{\otimes}

\usepackage{mathtools}

\newcommand\nounderline[1]{ #1} 
\newcommand\dolemma[1]{\vskip 5pt \noindent{\bf \underline{Lemma #1.}\ }}

\newcommand\dotheorem[1]{\vskip 5pt \noindent {\bf \underline{Theorem #1.}\ }}

\newcommand\doexample[1]{\vskip 5pt \noindent {\bf \underline{Example #1.}\ }}

\newcommand\doproof{\vskip 5pt \noindent{\bf \nounderline{Proof:}\ }}

\newcommand\tombstone{\rule{.6em}{.6em}}

\newcounter{numitem}
\newcommand{\numitem}[1]{\refstepcounter{numitem}\thenumitem\label{#1}}

\newcommand\bdy{\partial}

\newcommand\chainC{A}
\newcommand\chainD{B}
\newcommand\codec{a}
\newcommand\coded{b}

\newcommand\bdya{\bdy_a}
\newcommand\bdyb{\bdy_b}

\newcommand\domc{n_\codec}
\newcommand\codc{m_\codec}
\newcommand\domd{n_\coded}
\newcommand\codd{m_\coded}

\newcommand\Pauli{{\mathcal P}}
\newcommand\Cliff{{\mathcal C}}

\title{Limitations on transversal gates \\ for hypergraph product codes}

\author{Simon Burton}
\author{Dan Browne}
\affil{\small\it Department of Physics and Astronomy, University College London, London, WC1E 6BT, UK}
\date{\today}
\begin{document}
\maketitle

\begin{abstract}
We analyze the structure of the logical
operators from a class of quantum codes that generalizes the surface codes.
These codes are hypergraph product codes, restricted to the vertical sector.
By generalizing an argument of Bravyi and K{\"o}nig, we find that transversal
gates for these codes must be restricted to the Clifford group.
\end{abstract}

%
%

\section{Introduction}

Since the dawn of time humankind has striven to calculate and compute.
More recently we strive to build a quantum computer.
Like any computer, this has two ingredients: a reliable storage of information,
and the ability to manipulate this information. (Input/output comes later.)
These two ingredients are at odds with each other: one tries to prohibit
any change, and the other is all about change.
We seek a \emph{quantum code} that can do both.

The Kitaev code, which has the geometry of a torus, and its
flat cousin, called the surface code, are the leading contenders~\cite{Dennis2002}.
Industry has this goal within sight.
To reach universality with these codes we can do so via 
braiding~\cite{Fowler2012},
lattice surgery~\cite{Horsman2012,Fowler2018,Litinski2019},
or twists~\cite{Brown2017}.
Other approaches 
use a 3D architecture~\cite{Vasmer2019}, 
possibly with time~\cite{Bombin2018,Brown2020}
as one of the dimensions.
The problem with all these topological codes is that they have a vanishing rate:
as we build larger codes to decrease the error rate, the number of
logical qubits per physical qubit tends to zero.

Looking 
over the implementation horizon we find codes with vastly superior performance:
the family of hypergraph product codes~\cite{Tillich2014}. 
These still have a bounded number of interactions between physical qubits;
with the additional difficulty that these are non-local interactions.
The above story performed with the surface code has a parallel
story for hypergraph product codes, which are a direct generalization of
the surface code.
Good decoders exist \cite{Panteleev2019}, 
and we can braid punctures to perform Clifford gates \cite{Krishna2019}.

\begin{figure}[t]
\centering
\includegraphics[scale=1.0]{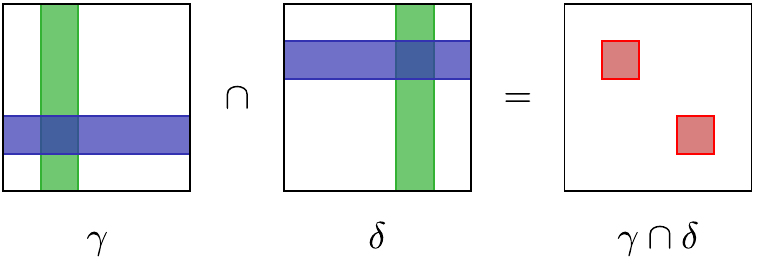}
\caption{The intersection of the support of two complete sets of logical 
operators is correctable under the hypothesis of Theorem \ref{th:main}.}
\label{fig:intersect}
\end{figure}

In this work we show that a family of hypergraph product codes are unable to achieve 
non-Clifford gates transversally.
This is a generalization of an argument of Bravyi and K{\"o}nig \cite{Bravyi2013}
which applies to surface/toric codes.
The Bravyi-K{\"o}nig argument relies on computing the intersections
of two sets of generic logical operators, and showing that this intersection is
correctable. See Figure~\ref{fig:intersect}.
For the surface code, or toric code, this intersection has size $O(1)$
and so cannot support any logical operator as the weight of these
scales with the lattice size.

For hypergraph product codes, the situation is more delicate.
In this case the intersection of generic logical operators has
support on an extensive set of qubits, and so we cannot rely
on constant distance bounds. 
This intersection region will have size $O(k^2)$,
where $k$ is the dimension of the underlying classical codespace.
At first glance this region would seem unlikely to be correctable,
as $k$ is comparable to the distance of the quantum hypergraph product code.
However, we show that under some mild assumptions on the
underlying classical codes (we call this assumption \emph{robustness})
this region is indeed correctable and so the Bravyi-K{\"o}nig argument
can be extended to these codes.
This is the main result of the paper, Theorem~\ref{th:main} below.

In summary, these hypergraph product codes may be excellent for
protecting quantum information, but there are restrictions on how 
this quantum information can then be manipulated. 
This work is the first result demonstrating such restrictions.

%
%

\section{Quantum stabilizer codes}

Fix a number of qubits $n$.
The Pauli group $\Pauli_n$ is generated by $n$-fold tensor products of 
the Pauli matrices 
$$
I_2=\left(\begin{array}{cc}1&0\\0&1\end{array}\right),\ \ 
X=\left(\begin{array}{cc}0&1\\1&0\end{array}\right),\ \ 
Z=\left(\begin{array}{cc}1&0\\0&-1\end{array}\right),\ \ 
Y=\left(\begin{array}{cc}0&-i\\i&0\end{array}\right),
$$
and phases $\{\pm 1, \pm i\}.$
Writing $V=\Complex^2$ for the single qubit Hilbert space,
the group $\Pauli_n$ acts tautologically on the $n$-qubit Hilbert space $V^{\tensor n}$
of dimension $2^n.$
Any subgroup $S \le \Pauli_n$ 
will \emph{fix} a subspace of $V^{\tensor n}$ pointwise:
$$
    \mathrm{Fix}_V^{\tensor n}(S) := \{ v\in V^{\tensor n}\ |\ gv = v,\  \forall g \in S \},
$$
and any subspace $W\le V^{\tensor n}$ will be \emph{stabilized} by a subgroup of $\Pauli_n$:
$$
    \mathrm{Stab}_{\Pauli_n}(W) := \{ g\in \Pauli_n\ |\ gw = w,\ \forall w \in W \}.
$$
An abelian subgroup $S\le \Pauli_n$ that does not contain $-I_2$ has
particularly nice structure. 
We call any such subgroup $S$ a \emph{stabilizer group}. 
Together with the tautological action this is known as a (quantum) \emph{stabilizer code}.
The subspace $\mathrm{Fix}_{V^{\tensor n}}(S)$ is the protected \emph{codespace}.

Operators $g\in\Pauli_n$ built as tensor product of $I_2$ and $X$ are called
$X$-type operators and denoted $\Pauli_n^X$.
Similarly, tensor products of $I_2$ and $Z$ are called $Z$-type operators and denoted $\Pauli_n^Z.$
If $g\in\Pauli_n$  commutes with every $h\in S$, then $G$  will preserve the
codespace, without needing to fix it pointwise. 
These are the \emph{logical operators} of the stabilizer code.
These logical operators are isomorphic to $\Pauli_k$, for some $k$, and so there are 
$k$ logical encoded qubits~\cite{Nielsen2010}.

When $S$ is generated by $X$-type or $Z$-type operators (inclusively)
we call this a \emph{CSS stabilizer code}~\cite{Gottesman1996,Calderbank1997}.
We state a simple lemma describing the logical operators of these codes.
Enumerating the $n$ qubits as $\{1,...,n\}$, we say that an operator $g\in\Pauli_n$ 
has \emph{support} on a subset $\gamma \subseteq \{1,...,n\}$
when the tensor factors of $g$ are equal to $I_2$ at indexes not in $\gamma.$

\dolemma{\numitem{lemma:css}} 
In a CSS stabilizer code, if a set of qubits $\gamma$
supports a non-trivial logical operator, then $\gamma$ supports either
a non-trivial $X$-type logical operator, or a non-trivial $Z$-type logical operator (or both).
\doproof 
An operator $g\in\Pauli_n$ commutes with every element of $S$ exactly when
it commutes with every generator of $S$.
\tombstone 

%
%

The $n$-qubit Clifford hierarchy~\cite{Gottesman1999} is a sequence $\{\Cliff_n^1, \Cliff_n^2, ...\}$
of sets of unitary operators, $\Cliff_n^l \subset \Unitary(V^{\tensor n})$.
These are defined inductively as $\Cliff_n^1:=\Pauli_n$, and 
$$
    \Cliff_n^{l+1} := \{ u\in\Unitary(V^{\tensor n})\ |\ u p u^{-1}\in\Cliff_n^l,\ \forall p\in\Pauli_n \}
$$
for $l\ge 1.$
We also define $\Cliff_n^0$ to be the phase group 
$\Cliff_n^0 = \{\pm I_2^{\tensor n}, \pm iI_2^{\tensor n}\}.$
The sets $\Cliff_n^0,\Cliff_n^1$ and $\Cliff_n^2$ are closed under inverse and
multiplication and so these are groups.
Specifically, $\Cliff_n^2$ is called the $n$-qubit \emph{Clifford group}.
For $l> 2$, $\Cliff_n^l$ is no longer a group~\cite{Zeng2008}.

\dolemma{\numitem{lemma:clifford}} 
A unitary $u\in\Unitary(V^{\tensor n})$ is in the $n$-qubit Clifford group exactly when 
$$
[upu^{-1}, q] = u p u^{-1} q u p^{-1} u^{-1} q^{-1} \in \Cliff_n^0
$$
for all $p,q\in\Pauli_n.$
\doproof 
The set of all unitaries 
$u\in\Unitary(V^{\tensor n})$ such that $upu^{-1}p^{-1}=\pm I$ for all $p\in\Pauli_n$,
equals the Pauli group.
Therefore the result follows from the definition of $\Cliff_n^2.$
\tombstone 

A \emph{transversal gate} for a stabilizer code $S\le\Pauli_n$ is an operator of
the form
$u = u_1\tensor ...\tensor u_n$
with each $u_i$ an arbitrary unitary on $V$,
such that $u$ commutes with every element of $S$.
Just like the logical operators of the code, these transversal gates also
preserve the codespace, without needing to fix it pointwise.
The set of all such transversal gates is a subgroup of $\Unitary(V)^{\tensor n}.$

For any operator $u\in \Unitary(V)^{\tensor n}$
the \emph{support} of $u$ is the set of qubits, or indices,
on which $u$ acts non-trivially.
Given a stabilizer code $S$, 
a subset of the qubits 
is called \emph{correctable}
when the only logical operators supported on the subset are in $\Cliff_n^0$.
Any logical operator in $\Cliff_n^0$ will be called \emph{trivial logical operator}.

The central idea behind the proof of Theorem~\ref{th:main} is
summarized in the following result.

\dotheorem{\numitem{th:clifford}}(Bravyi-K{\"o}nig argument \cite{Bravyi2013}.) 
Let $S$ be a stabilizer code such that the intersection of the support of any
two logical operators is a correctable set of qubits.
Then the action of any 
transversal gate for $S$ on the $k$ logical qubits is an element of
the Clifford group $\Cliff_k^2.$
\doproof 
This is a consequence of the previous lemma.
Let $u\in\Unitary(V)^{\tensor n}$ be a transversal gate for $S$.
This operator restricts to a unitary on the logical qubits
$\tilde{u}\in\Unitary(V^{\tensor k})$.
Similarly for logical operators $p,q\in\Pauli_n$ and
restriction $\tilde{p},\tilde{q}\in\Pauli_k$.
Then 
$[upu^{-1}, q]\in\Unitary(V)^{\tensor n}$ 
is non-trivial only on the intersection of the support of $p$ and $q$.
Because this region is correctable, we have $[upu^{-1}, q]\in \Cliff_n^0,$
which implies  $[\tilde{u}\tilde{p}\tilde{u}^{-1}, \tilde{q}]\in \Cliff_k^0.$
\tombstone 

The groups $\Pauli_n^X$ and $\Pauli_n^Z$ are abelian, and moreover, these groups are 
isomorphic to the additive group of the $\Field_2$-linear vector
space $\Field_2^n.$
In the remainder of this work, we switch to this additive group notation.
Products of operators are computed as sums of $\Field_2$ vectors.
Commutators of $X$-type and $Z$-type operators are calculated via the evident 
$\Field_2$ inner product.
For $u\in\Field_2^n$ an $X$-type operator, and $v\in\Field_2^n$ a $Z$-type
operator, this is just $u^\top v\in\Field_2.$

%
%

\section{Linear codes}\label{sec:codes}

We work using linear vector spaces over the field with
two elements $\Field_2.$
Such vector spaces are constructed as $\Field_2^n$ for
some number $n$.
We purposely confuse the distinction and notation between the vector
space $\Field_2^n$, the dimension $n$,
and the $n$-element set of basis vectors implied.
Specifically, letters $n,m$ and $k$ below, possibly with subscript,
may denote any of these.
It should be clear from context which is meant.

A linear map $n\xrightarrow{\bdy}m$ has a {\it kernel}
which records dependencies among the columns of $\bdy$.
This is a linear map $k\xrightarrow{\Kernel(\bdy)}n$ such that $\bdy\cdot\Kernel(\bdy) = 0.$
Furthermore, this map is characterized by a universal property:
any other linear map $\bullet\xrightarrow{f}n$ with $\bdy f=0$
factors uniquely through $\Kernel(\bdy).$

Similarly, the {\it cokernel} records dependencies among the rows of $\bdy$.
This is a linear map $n\xrightarrow{\Cokernel(\bdy)}k^{\top}$
such that $\Cokernel(\bdy)\cdot \bdy = 0.$
(Here, the symbol ${}^\top$ in the notation $k^\top$ is a formal decoration on the symbol $k$.)
Any other linear map $m\xrightarrow{g}\bullet$ with $g\bdy=0$
factors uniquely through $\Cokernel(\bdy).$
Evidently we have the relation
$ \Kernel(\bdy)^{\top} = \Cokernel(\bdy^{\top}).$

Using both the kernel and the cokernel
we can construct the following sequence of linear maps:
$$
0\to k\xrightarrow{\Kernel(\bdy)}n\xrightarrow{\bdy}m\xrightarrow{\Cokernel(\bdy)}k^\top\to 0.
$$
This sequence is exact, so we have the fundamental identity 
\begin{align}\label{eq:euler}
k-n+m-k^\top=0.
\end{align}
This is the Euler characteristic of this sequence.
(See the book \cite{Ghrist2014}, in particular section 5.9, for 
a user-friendly introduction to sequences, homology and the Euler characteristic.)

The \emph{span} of $n\xrightarrow{\bdy}m$ 
is a linear map $m-k^\top\xrightarrow{\Span(\bdy)}m$ that factors
through $\bdy$. This map is also characterized by a universal property:
any other linear map with codomain $m$ that factors through $\bdy$ 
factors uniquely through $\Span(\bdy).$
Dually, the \emph{cospan} of $n\xrightarrow{\bdy}m$ 
is a linear map $n\xrightarrow{\Cospan(\bdy)}n-k$
that factors through $\bdy$ (on the other side). The universal property
characterizing the cospan is that any other linear map with domain $n$ factoring through $\bdy$
will factor uniquely through $\Cospan(\bdy).$
From (\ref{eq:euler}) we find that $m-k^\top=n-k$ which is the familiar fact that
the column-span and row-span of a matrix have the same dimension.

A classical linear code is usually defined to be a subspace of some
finite dimensional vector space $\Field_2^n.$
We are taking a relentlessly ``active'' approach to linear algebra:
the maps are primary, not the spaces.
For our purposes, a \emph{classical linear code} is a linear map $n\xrightarrow{\bdy}m$.
The \emph{codespace} is the kernel of this map.
We will avoid the temptation to define the cocodespace as the cokernel.
We also call $n\xrightarrow{\bdy}m$ a \emph{parity check matrix},
and $\Kernel(\bdy)^\top$ the \emph{generator matrix} for this code.

A \emph{quantum code} is defined to be a pair of linear maps
$$
    m_Z\xrightarrow{H_Z^\top} n \xrightarrow{H_X} m_X
$$
such that $H_X H_Z^\top = 0.$
From this identity it follows we can take the quotients
\begin{align*}
    L_Z^\top = \Kernel(H_X) / \Span(H_Z^\top),\ \  
    L_X = \Cokernel(H_Z^\top) / \Cospan(H_X)
\end{align*}
which we call the \emph{logical operators} for the code.
We will use the following block matrix notation for these quotients:
$$
L_Z^{\top} = \left(
    \begin{array}{c;{2pt/2pt}c}
        \Kernel(H_X) & H_Z^\top 
    \end{array}
\right), \ \ 
L_X = \left(\begin{array}{c}
\Cokernel(H_Z^\top) \\
\hdashline[2pt/2pt]
H_X 
\end{array}\right),
$$
where the dashed line indicates the modulo (quotient).

%
%

\section{Punctures}

We write the generic set of size $j$ as $[j]$,
which we also identify with the canonical basis of $\Field_2^j.$
Given a subset $\gamma\subset [j]$, we say a vector $v\in\Field_2^j$
has {\it support on} $\gamma$ if the non-zero components of $v$
form a subset of $\gamma$, in symbols: $\{i | v_i\ne 0\}\subset \gamma.$

A linear map $G:n\to k$ is called $e$-\emph{puncturable},
for some natural number $e$, if there  exists a subset $\gamma\subset [n]$ of
size $e$ such that no non-zero vector in the cospan of $G$ 
has support on $\gamma$.
In terms of the matrix for $G$ this corresponds to
being able to delete the columns indexed by $\gamma$ while
maintaining the rank of $G$.
We call such a set $\gamma$ an {\it $e$-puncture} of $G$,
or a {\it puncture} of $G$ when $e$ is understood.
When $G$ is the generator matrix of a classical code
we recover the notion of puncturing a classical code ($e$ times) well known 
in the literature~\cite{Macwilliams1977}.
However, we also apply this definition to arbitrary
linear maps.

Going further, a linear map $G:n\to k$ is called
$e$-{\it bipuncturable}, for some natural number $e$, 
if there exists two disjoint subsets $\gamma\subset [n], \delta\subset [n]$,
such that $\gamma$ is an $e$-puncture of $G$ and
$\delta$ is an $e$-puncture of $G$.
We call such a pair of sets $(\gamma,\delta)$ an 
{\it $e$-bipuncture} of $G$
or a {\it bipuncture} of $G$ when $e$ is understood.

\doexample{\numitem{ex:puncture}}
The repetition code $4\xrightarrow{\bdy} 3$ is given by the parity check
matrix 
$$
    \bdy = 
\left(\begin{array}{cccc}
1 & 1 & . & . \\
. & 1 & 1 & . \\
. & . & 1 & 1 \\
\end{array}\right).$$
This linear map is only 1-puncturable because it contains
all weight two vectors in its cospan.
The generator matrix for this code is given by 
$ G = \Kernel(\bdy)^\top = (1 1 1 1),$
and we can choose any three element subset of $[n=4]$ as a puncture of $G$.
Moreover, we can choose any two element subset $\gamma\subset[4]$
with $\delta=[4]-\gamma$, to demonstrate that $G$ is 2-bipuncturable.
\tombstone 

More generally,
given a classical code with distance $d$,
any subset of $[n]$ with size less than $d$ serves as a puncture
for the generating matrix,
and so this matrix is $(d-1)$-puncturable. 

\dolemma{\numitem{lemma:puncture}} 
Any linear map $\bdy:n\to m$ with $\Kernel(\bdy):k\to n$,
is $k$-puncturable.
\doproof
We can row-reduce $\bdy$ without changing any of the punctures.
By reordering the columns, the row-reduced $\bdy$ has block matrix form
$$ \bdy = \left(\begin{array}{cc} I_{n-k} & J \\ 0 & 0 \end{array}\right) $$
where $J$ is an arbitrary matrix with $k$ columns.
Puncture these $k$ columns.
\tombstone

%
%
\section{Hypergraph product codes}

Given classical codes $\chainC = \{\domc\xrightarrow{\bdya} \codc \}$
and $\chainD = \{\domd\xrightarrow{\bdyb} \codd \}$
we define a quantum code $\chainC\tensor \chainD$ as:
\begin{align}\label{eq:quantum}
    \domc\tensor \domd \xrightarrow{H_Z^{\top}} 
    \domc \tensor \codd\oplus \codc \tensor \domd \xrightarrow{H_X}
    \codc\tensor \codd.
\end{align}
Using block matrix notation, 
\begin{align}\label{eq:product}
H_Z^{\top} = \left( \begin{array}{c} I_{\domc}\tensor \bdyb \\ \bdya\tensor I_{\domd} \end{array} \right),
\ \ \ 
H_X = \big( \bdya\tensor I_{\codd}\ \ \  I_{\codc}\tensor \bdyb \big).
\end{align}
A quick calculation shows that $H_X H_Z^\top = 2\bdya\tensor\bdyb = 0$
thanks to $\Field_2$ arithmetic, and so this is indeed a quantum code.
This \emph{hypergraph product code} $A\tensor B$ can also be notated as the following diagram
\begin{center}
\includegraphics[]{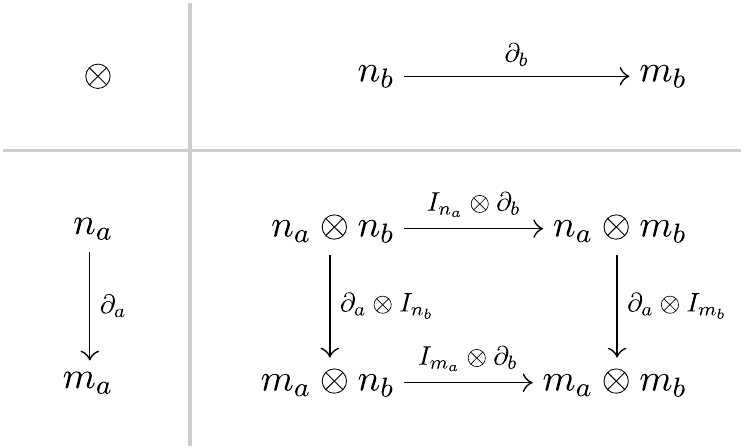}
\end{center}
where the inner square is not commutative in general.
We are using the tensor symbol for the hypergraph product $A\tensor B$,
because this is indeed the tensor product of $A$ and $B$ 
as chain complexes. This product is also known as the \emph{homological product}.

\dolemma{\numitem{lemma:euler}} 
With the above notation, the quantum code $\chainC\tensor \chainD$
has $k$ logical qubits given by
$ k = k_\codec k_\coded^\top + k_\codec^\top k_\coded. $
\doproof 
By inspecting (\ref{eq:product})
we find linear maps:
    $$k_\codec\tensor k_\coded\xrightarrow{\Kernel(H_Z^\top)} \domc\tensor\domd$$
and
    $$\codc\tensor\codd \xrightarrow{\Cokernel(H_X)} k_\codec^\top\tensor k_\coded^\top.$$
Now use this to 
extend the sequence (\ref{eq:quantum}):
$$
    0\to k_\codec\tensor k_\coded\xrightarrow{\Kernel(H_Z^\top)}
    \domc\tensor \domd \xrightarrow{H_Z^{\top}} 
    \domc \tensor \codd\oplus \codc \tensor \domd \xrightarrow{H_X}
    \codc\tensor \codd
    \xrightarrow{\Cokernel(H_X)} k_\codec^\top\tensor k_\coded^\top\to 0.
$$
This sequence is exact everywhere except in the middle where the
homology is $k$ (dimensional).
Therefore, the Euler character is:
\begin{align*}
    k 
    &= k_\codec k_\coded - n_\codec n_\coded + n_\codec m_\coded + m_\codec n_\coded - m_\codec m_\coded + k_\codec^\top k_\coded^\top \\
    &=  k_\codec k_\coded + (n_\codec-m_\codec)(m_\coded-n_\coded) + k_\codec^\top k_\coded^\top \\
    &=  k_\codec k_\coded + (k_\codec-k_\codec^\top)(k_\coded^\top-k_\coded) + k_\codec^\top k_\coded^\top 
    \ \ \ \ \text{using (\ref{eq:euler})}\\
    &= k_\codec k_\coded^\top + k_\codec^\top k_\coded. \ \  \tombstone
\end{align*} 

The qubits (vectors) in $\domc\tensor \codd$ are
called {\it vertical} qubits, and the qubits in 
$\codc\tensor \domd$ are called {\it horizontal} qubits.
This convention is motivated by the following example.

\doexample{\numitem{ex:surface}} (Surface code). 
Continuing to use the above notation, we take 
$\chainC = \{\domc\xrightarrow{\bdya} \codc \}$
to be a repetition code and 
$\chainD = \{\domd\xrightarrow{\bdyb} \codd \}$
to be the dual (transpose) code:
$$
    \bdya = 
\left(\begin{array}{cccc}
1 & 1 & . & . \\
. & 1 & 1 & . \\
. & . & 1 & 1 \\
\end{array}\right),\ \ \ 
\bdyb = 
\left(\begin{array}{ccc}
1 & . & . \\
1 & 1 & . \\
. & 1 & 1 \\
. & . & 1 \\
\end{array}\right).
$$
If we use the number of dimensions as shorthand for the
$\Field_2$ vector space itself, we have $\bdya:4\to 3$ and $\bdyb:3\to 4$,
and also
$$
\begin{array}{rr}
k_\codec = 1, & k_\coded = 0, \\
k_\codec^\top = 0, & k_\coded^\top = 1.\\
\end{array}
$$
The quantum code $\chainC\tensor \chainD$ is then:
$$
    4\times 3 \xrightarrow{H_Z^{\top}} 3\times 3 + 4\times 4 
        \xrightarrow{H_X} 3\times 4.
$$
Pictorially this corresponds to $4\times 3$ faces,
$3\times 3$ horizontal edges, $4\times 4$ vertical edges,
and $3\times 4$ vertices:
\begin{center}
\includegraphics[scale=0.8]{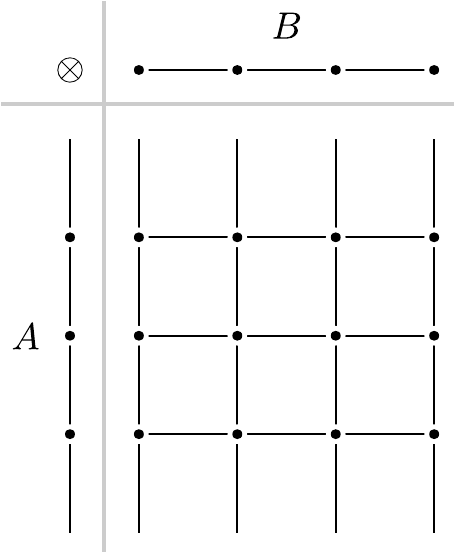}
\end{center}
\tombstone 

\begin{figure} 
\centering
    \begin{subfigure}[t]{0.45\textwidth}
        \includegraphics[scale=1.0]{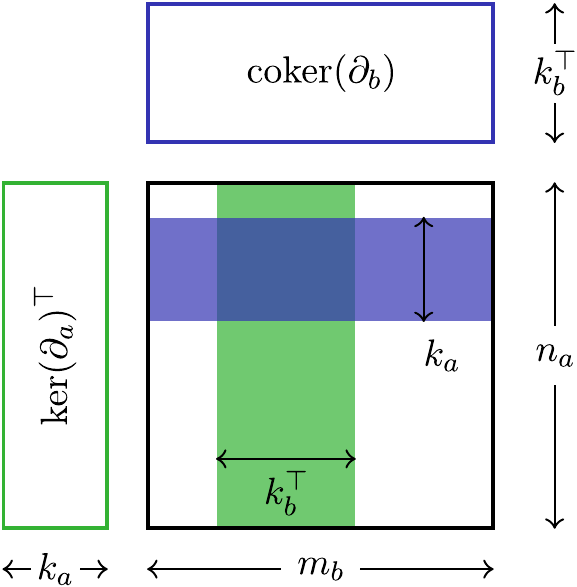}
\caption{
Vertical sector with $n_a m_b$ many qubits,
and matrices for $\Kernel(\bdya)^\top$ and $\Cokernel(\bdyb)$.
}
        \label{fig:homproda}
    \end{subfigure}\ \ \ \ \ \ \ 
    \begin{subfigure}[t]{0.45\textwidth}
        \ \ \ \ \ \includegraphics[scale=1.0]{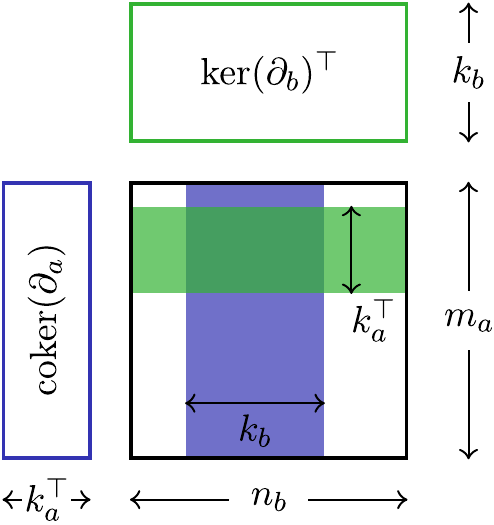}
\caption{%
Horizontal sector with $m_a n_b$ many qubits,
and matrices for $\Cokernel(\bdya)$ and $\Kernel(\bdyb)^\top$.
}
        \label{fig:homprodb}
    \end{subfigure}
\caption{
Illustrating the construction of the logical operators 
in the hypergraph product code,
from the underlying classical codes.
(i) 
A complete set of $k_ak_b^\top$ many purple $L_X$ operators is supported on $k_a$ many horizontal
strips, each of which supports $k_b^\top$ many operators from the
rows of $\Cokernel(\bdy_b).$
Similarly, we have $k_b^\top k_a$ many green $L_Z$ operators.
(ii) The horizontal qubits support $k_b k_a^\top$ many purple $L_X$,
and $k_a^\top k_b$ many green $L_Z$ operators.
}
    \label{fig:homprod}
\end{figure} 

The next theorem describes the structure of the
logical operators as illustrated in Figure~\ref{fig:homprod}.
It is a concrete restatement 
of the formula $k=k_ak_b^\top+k_a^\top k_b$ from the previous lemma.
The standard basis for $\domc$ we write as $[\domc]$, similarly for
$\codc,\codd$ and $\domd.$ Notation such as $[\domc]\times[\codd]$ 
denotes the basis for $\domc\tensor \codd$, which is the
space for the vertical qubits.

\dotheorem{\numitem{th:logops}} 
Let $A\tensor B$ be a hypergraph product code as above.
A complete set of $L_Z$ operators is supported on the qubits:  

\begin{tabular}{ l l }
($L_Z$-vertical)  & 
$[\domc]\times\gamma_{Zv}$ where $\gamma_{Zv}$ is any 
    $k_\coded^\top$-puncture of $\bdyb^\top$, and \\
($L_Z$-horizontal)  & 
$\gamma_{Zh}\times[n_b]$ where $\gamma_{Zh}$ is any 
    $k_a^\top$-puncture of $\bdya^\top$. \\
\end{tabular}

A complete set of $L_X$ operators is supported on the qubits: 

\begin{tabular}{ l l }
($L_X$-vertical) & 
$\gamma_{Xv}\times [\codd]$ where $\gamma_{Xv}$ is any $k_\codec$-puncture of $\bdya$, and \\
($L_X$-horizontal) & 
$[m_a]\times\gamma_{Xh}$ where $\gamma_{Xh}$ is any $k_\coded$-puncture of $\bdyb$.
\end{tabular}

Specifically, we have the following block matrix form for
the logical operators
$$
L_Z^\top = \left(
    \begin{array}{cc;{2pt/2pt}c}
        \Kernel(\bdya)\tensor \gamma_{Zv}^\top & 0 & I_{n_a}\tensor\bdyb \\
        0 & \gamma_{Zh}^\top \tensor \Kernel(\bdyb) & \bdya\tensor I_{n_b} 
    \end{array}
\right),
$$
where the basis for the rows is $v=n_a\tensor m_b$ and $h=m_a\tensor n_b$, 
and
$$
L_X = \left(\begin{array}{cc}
\gamma_{Xv} \tensor\Cokernel(\bdyb) & 0 \\
0 & \Cokernel(\bdya) \tensor \gamma_{Xh}  \\
\hdashline[2pt/2pt]
\bdya\tensor I_{m_b} & I_{m_a}\tensor \bdyb 
\end{array}\right),
$$ 
where the basis for the columns is $v$ and $h.$
In this notation, a puncture $\gamma\subseteq [n]$ is promoted to a 
$|\gamma|\times n$ matrix by taking
the rows of the identity matrix $I_n$ indexed by $\gamma.$
The dashed line separates ``generators'' from ``relations'', as in Section~\ref{sec:codes}.
\doproof 
We claim that the kernel of $H_X$ is spanned by the columns of a matrix as:
$$
\Kernel(H_X) = \Span
\left(\begin{array}{ccc}
\Kernel(\bdya)\tensor I_{m_b} & 0 & I_{n_a}\tensor \bdyb \\
0 & I_{m_a}\tensor \Kernel(\bdyb) &  \bdya\tensor I_{n_b}
\end{array}\right),
$$
and the cokernel of $H_Z^{\top}$ 
is spanned by the rows of a matrix as:
$$
\Cokernel(H_Z^{\top}) = \Cospan
\left(\begin{array}{cc}
I_{n_a}\tensor\Cokernel(\bdyb) & 0 \\
0 & \Cokernel(\bdya)\tensor I_{n_b} \\
\bdya\tensor I_{m_b} & I_{m_a}\tensor \bdyb 
\end{array}\right).
$$
That these matrices lie within $\Kernel(H_X)$ and $\Cokernel(H_Z^\top)$
respectively can be seen by composing with $H_X, H_Z^\top$ respectively.
We next count the independent logical operators and find 
$ k_a k_b^\top + k_a^\top k_b$ many,
then by Lemma~\ref{lemma:euler} we will have found all of them.

Let $\gamma_{Zv}$ be any $k_b^\top$-puncture of $\bdy_b^\top$.
This puncture always exists by Lemma~\ref{lemma:puncture}.
Any solution of 
$$
\left(\begin{array}{ccc}
\Kernel(\bdya)\tensor \gamma_{Zv}^\top & 0 & I_{n_a}\tensor \bdyb \\
0 & I_{m_a}\tensor \Kernel(\bdyb) &  \bdya\tensor I_{n_b}
\end{array}\right)
\left(\begin{array}{c}x\\y\\z\end{array}\right) = \left(\begin{array}{c}0\\0\end{array}\right)
$$
will have $x=0$. 
Counting the columns of $\Kernel(\bdy_a)\tensor\gamma_{Zv}$ we find $k_a k_b^\top.$
Therefore, we have found $k_a k_b^\top$ linearly independent $L_Z$ logical operators:
$$
\left(
    \begin{array}{c;{2pt/2pt}c}
        \Kernel(\bdya)\tensor \gamma_{Zv}^\top & I_{n_a}\tensor\bdyb \\
        0 & \bdya\tensor I_{n_b} 
    \end{array}
\right).
$$
The other $k_a^\top k_b$ operators for $L_Z$  are found similarly,
and also for the $L_Z$ operators.
\tombstone 

The form of the logical operators in the above theorem has a particular
structure which we will use repeatedly below.
An $X$-type or $Z$-type logical operator is called {\it taut}
when it is ``as straight and thin as possible''.
Specifically,
taut $L_Z$-vertical logical operators are of the form
$v\tensor j$ with $v\in\Cospan(\Kernel(\bdya)^\top)$
and $j\in [\codd]$ a basis vector, and taut 
$L_X$-vertical logical operators of the form
$i\tensor u$ with $u\in\Cospan(\Cokernel(\bdyb))$
and $i\in [\domc]$ a basis vector.
See Figure \ref{fig:homproda}.
A similar definition holds for horizontal taut logical operators.
Note that taut operators are not necessarily of minimal weight.

\doexample{\numitem{ex:surfctd}} (Surface code continued). 
For the $L_Z$-vertical logical operators we compute $\Kernel(\bdya)\tensor\gamma_{Zv}^\top$:
$$
\Kernel(\bdya)=\left(\begin{array}{c}1\\1\\1\\1\end{array}\right)
$$
and we take $\gamma_{Zv}$ to be a $1\times 4$ matrix that is non-zero at any one
coordinate, to get a 1-puncture of $\bdyb^\top$.
For the $L_X$-vertical logical operators we compute 
$\gamma_{Xv}\tensor\Cokernel(\bdyb)$.
In this case we have $\Cokernel(\bdyb)=(1 1 1 1)$
and $\gamma_{Xv}$ is a $1\times 4$ matrix non-zero at any one coordinate.

The $L_Z$-horizontal logical operators are trivial because $k_a^\top=0$, and
the $L_X$-horizontal logical operators are trivial because $k_b=0.$
\tombstone

%
%

\section{Main result}

We say that a hypergraph product code $\chainC\tensor \chainD$ is 
{\it restricted to the vertical sector}
when 
$\Kernel(\bdyb)$ is trivial and/or 
$\Cokernel(\bdya)$ is trivial.
See Figure \ref{fig:homprod}.
This happens precisely when $k_a^\top k_b=0.$
The surface code of Example \ref{ex:surface} is one such code.
The toric code is a counter-example, in this case $k_a = k_a^\top = k_b = k_b^\top = 1.$
The next lemma shows how to restrict Theorem~\ref{th:logops} to the
vertical sector.

\dolemma{\numitem{lemma:complete}} 
Given a hypergraph product code $\chainC\tensor \chainD$ restricted 
to the vertical sector, we have 
(i) a complete set of $L_Z$ operators is supported
on $[\domc]\times\gamma_Z$ where $\gamma_Z$ is any
$k_\coded^\top$-puncture of $\bdyb^\top$, and
(ii) a complete set of $L_X$ operators is supported
on $\gamma_X\times [\codd]$ where $\gamma_X$ is any
$k_\codec$-puncture of $\bdya$. 
\doproof 
We use Theorem~\ref{th:logops} and assume $k_a^\top k_b=0$. 
If $k_b=0$ then $\Kernel(\bdyb)=0$
so $L_Z$-horizontal is trivial,
and any $k_b$-puncture of $\bdyb$ is empty so $L_X$-horizontal is trivial.
Similarly, if $k_a^\top=0$ then $\Cokernel(\bdya)$ is trivial
so $L_X$-horizontal is trivial,
and any $k_a^\top$-puncture of $\bdya^\top$ is empty so $L_Z$-horizontal is trivial.
The remaining operators are $L_Z$-vertical and $L_X$-vertical, as required.
\tombstone 

In Theorem~\ref{th:logops} we showed how representatives of the logical
operators can be chosen to be taut.
The next result shows how more general logical operators can be decomposed into
taut logical operators, when restricted to the vertical sector.

\dolemma{\numitem{lemma:taut}} 
Given a hypergraph product code $\chainC\tensor \chainD$ restricted 
to the vertical sector,
any $Z$-type (resp. $X$-type) logical operator supported on the vertical qubits
is a product of disjoint taut $Z$-type (resp. $X$-type) logical operators.
\doproof 
From the form of the logical operators in Theorem~\ref{th:logops},
we have the block matrix 
$$
L_Z^{\top} = \left(
    \begin{array}{cc;{2pt/2pt}c}
        \Kernel(\bdya)\tensor I_{m_b} & 0 & I_{n_a}\tensor\bdyb \\
        0 & I_{m_a}\tensor \Kernel(\bdyb) & \bdya\tensor I_{n_b} 
    \end{array}
\right).
$$ 
This notation is to be read as ``column vectors modulo column vectors''
with the dashed line acting as the modulo.
Assuming the hypothesis we have that $k_a^\top k_b=0$ which
implies, also from Theorem~\ref{th:logops}, that the second column is trivial modulo the third column.
This gives
$$
L_Z^{\top} = \left(
    \begin{array}{c;{2pt/2pt}c}
        \Kernel(\bdya)\tensor I_{m_b} & I_{n_a}\tensor\bdyb \\
        0 & \bdya\tensor I_{n_b} 
    \end{array}
\right).
$$ 
An arbitrary logical Z-type operator $l_Z^\top$ is got from the span of this
matrix, which by the hypothesis must be zero on the bottom row:
$$
l_Z^\top = L_Z^\top \left(\begin{array}{c}x\\ y\end{array}\right) = 
 \left(\begin{array}{c}z \\ 0\end{array}\right),
$$
for some matrices $x, y, z$.
Solving this equation in generality, we require $(\bdya\tensor I_{n_b})y = 0$
which gives $y=\Kernel(\bdya)\tensor I_{n_b}.$
Expanding out, we get:
\begin{align*}
z &= (\Kernel(\bdya)\tensor I_{m_b})x + \Kernel(\bdya)\tensor \bdyb \\
    &=  (\Kernel(\bdya)\tensor I_{m_b})x + (\Kernel(\bdya)\tensor I_{m_b})(I_{k_a}\tensor \bdyb) \\
    &=  (\Kernel(\bdya)\tensor I_{m_b})(x + I_{k_a}\tensor \bdyb).
\end{align*}
This means that $l_Z^\top$ is got from the span of 
$$
\left(
    \begin{array}{c}
        \Kernel(\bdya)\tensor I_{m_b} \\
        0 
    \end{array}
\right)
$$ 
which is precisely the form of disjoint $Z$-type operators we seek.
The same calculation transposed shows the result for $X$-type logical operators:
$$
L_X = \left(\begin{array}{cc}
I_{n_a}\tensor\Cokernel(\bdyb) & 0 \\
0 & \Cokernel(\bdya)\tensor I_{n_b} \\
\hdashline[2pt/2pt]
\bdya\tensor I_{m_b} & I_{m_a}\tensor \bdyb 
\end{array}\right)
= \left(\begin{array}{cc}
I_{n_a}\tensor\Cokernel(\bdyb) & 0 \\
\hdashline[2pt/2pt]
\bdya\tensor I_{m_b} & I_{m_a}\tensor \bdyb 
\end{array}\right),
$$
and so on.
\tombstone 

Two regions (subsets) 
$\alpha,\beta\subset [\domc]\times[\codd]$
are called {\it horizontally separated} 
when for any $j\in[\codd]$ we have that $[\domc]\times\{j\}$
intersects at most one of $\alpha$ or $\beta$.
Similarly, $\alpha$ and $\beta$ are called {\it vertically separated}
when for any $i\in[\domc]$ we have that $\{i\}\times[\codd]$
intersects at most one of $\alpha$ or $\beta$.

\dolemma{\numitem{lemma:union}} (Union Lemma.) 
Consider a hypergraph product code $\chainC\tensor \chainD$ restricted 
to the vertical sector, with two regions 
$\alpha,\beta\subset [\domc]\times[\codd]$
that are separated horizontally and vertically.
If there exists a non-trivial logical operator supported on $\alpha\cup\beta$
then there exists a non-trivial logical operator supported on $\alpha$, or
there exists a non-trivial logical operator supported on $\beta$ (inclusive or).
\doproof 
By Lemma~\ref{lemma:css} we need only consider $X$-type and $Z$-type operators.
Suppose that $\alpha\cup\beta$ supports a non-trivial $X$-type logical operator.
Then by the previous lemma, $\alpha\cup\beta$ must support a non-trivial taut $X$-type operator.
This operator is supported on 
$\{i\}\times[\codd]$ for some $i\in[\domc]$, which intersects at most one
of $\alpha$ or $\beta$ because these are vertically separated.
Therefore we must have that at most one of 
$\alpha$ or $\beta$ supports this taut $X$-type logical operator.
Using the
horizontal separation of $\alpha$ and $\beta$
a similar argument applies to the $Z$-type logical operators.
\tombstone 

We call a classical code 
$\{n\xrightarrow{\bdy} m \}$
{\it robust} 
if the generator matrix $\Kernel(\bdy)^\top$ and the parity check matrix $\bdy$ are 
simultaneously $k$-bipuncturable, where $k$ is the dimension of the codespace.

\dotheorem{\numitem{th:main}}
Let 
$\chainC = \{\domc\xrightarrow{\bdya} \codc \}$ and 
$\chainD = \{\domd\xrightarrow{\bdyb} \codd \}$
be classical codes 
such that $\chainC\tensor\chainD$ is restricted to the vertical sector.
If $\chainC$ is robust and 
$\chainD^\top =  \{\codd\xrightarrow{\bdyb^\top} \domd \}$ 
is robust,
then any transversal gate for $\chainC\tensor \chainD$ is restricted to the Clifford group $\Cliff^2$.
\doproof 
We assume the hypothesis of the theorem.
Let $(\gamma_Z,\delta_Z)$ be a simultaneous $k_\coded^\top$-bipuncture of $\bdyb^\top$ and $\Kernel(\bdyb^\top)^\top = \mathrm{coker}(\bdyb)$.
Let $(\gamma_X,\delta_X)$ be a simultaneous $k_\codec$-bipuncture of $\bdya$ and $\mathrm{ker}(\bdya)^\top$.
From Lemma~\ref{lemma:complete}
we have a complete set of $L_Z$ operators supported on 
$[\domc]\times\gamma_Z$
and a complete set of $L_Z$ operators supported on 
$[\domc]\times\delta_Z.$
Similarly for $L_X$ operators supported on
$\gamma_X\times[\codd]$ and $\delta_X\times[\codd]$.
Combining these, we find that 
$\gamma:=\gamma_X\times[\codd] \cup [\domc]\times\gamma_Z$
supports a complete set of logical operators, as does
$\delta:=\delta_X\times[\codd] \cup [\domc]\times\delta_Z.$
See Figure~\ref{fig:intersect}.
We now claim that the intersection 
of these two sets $\gamma\cap\delta$
is correctable, that is, does not support any non-trivial
logical operator. 
The result then follows from Theorem~\ref{th:clifford}.

Using the disjointness of $\gamma_X$ and $\delta_X$, and 
the disjointness of  $\gamma_Z$ and $\delta_Z$ we find
\begin{align*}
\gamma\cap\delta 
&=
(\gamma_X\times[\codd] \cup [\domc]\times\gamma_Z )\cap(\delta_X\times[\codd] \cup [\domc]\times\delta_Z ) \\
&= 
(\gamma_X\times[\codd] \cap [\domc]\times\delta_Z) 
    \cup 
([\domc]\times\gamma_Z \cap \delta_X\times[\codd])
\end{align*}
and the RHS then satisfies the hypothesis of Lemma~\ref{lemma:union}.
Therefore we need only show the correctability of the two pieces,
$\alpha:=\gamma_X\times[\codd] \cap [\domc]\times\delta_Z$
and 
$\beta:=[\domc]\times\gamma_Z \cap \delta_X\times[\codd]$.

From the hypothesis, $\delta_Z$ punctures $\Cokernel(\bdyb)$ 
and so $\alpha$ cannot support any taut $L_X$ logical operator,
and by Lemma~\ref{lemma:taut} cannot support arbitrary $L_X$ logical operator.
Also, $\gamma_X$ punctures $\Kernel(\bdya)^\top$ and so
$\alpha$ cannot support any taut $L_Z$ logical operator,
and by Lemma~\ref{lemma:taut} cannot support arbitrary $L_Z$ logical operator.
By Lemma~\ref{lemma:css} we find that $\alpha$ cannot support
any logical operator, and so $\alpha$ is correctable.
A similar argument applies to show that $\beta$
is correctable.
\tombstone 

By swapping the factors $A\tensor B$ to $B\tensor A$ we see that this theorem 
also applies with ``vertical sector'' replaced with ``horizontal sector''.

In the next section we show that some mild conditions on
a classical code will guarantee it is robust.

%
%

\section{Characterizing robustness}

In this section we take $H=\bdy$ to be a parity check matrix
that has full rank, without loss of generality.
Adding linear dependant rows to $H$ does not change the punctures on $H$.
We also write $G=\Kernel(H)^\top$ for the generator matrix.

\dolemma{\numitem{lemma:normal}} (Canonical form). 
Any $k$-dimensional code with length $n$
has canonical form for generator matrix $G$ and parity
check matrix $H$,
\begin{align*}
    G &= (I_k\ \ \ J) \\
    H &= (J^\top\ \ I_m)
\end{align*}
up to reordering of columns, with $m=n-k.$
\doproof 
Row-reduction and column reordering  gives the form
\begin{align*}
    G &= (I_k\ \ \ J) \\
    H &= (K\ \ I_m).
\end{align*}
Using $HG^\top = 0$ we get $K = J^\top.$
\tombstone 

Writing such a code in this form, the set 
of indices $\{1,...,k\}$ are called 
{\it pivots} and the set of indices $\{k+1,...,n\}$ are called {\it copivots.}
We also use this terminology for matrices in arbitrary
row-reduced form (not necessarily column re-ordered to make the pivots sequential).

Next, we show how to puncture the copivots.

\dolemma{\numitem{lemma:copivot}} (Copivot Lemma.) 
With $G$ written in canonical form $G = (I_k\ \ J)$
any set of indices $\gamma\subset\{k+1,...,n\}$ punctures $G$.
\doproof 
For $v\in\Field_2^k$ any non-zero vector, $vG$ has non-zero
components on $\{1,...,k\}$ and so cannot be supported on $\gamma.$
\tombstone 

Slightly more tricky is puncturing the pivots.

\dolemma{\numitem{lemma:pivot}} (Pivot Lemma.) 
With $G$ written in canonical form $G = (I_k\ \ J)$
a set of indices $\gamma\subset\{1,...,k\}$ punctures $G$
if and only if $\gamma$ punctures $\Cokernel(J).$
\doproof 
The only way $\gamma\subset\{1,...,k\}$ can support a non-zero
vector in the cospan of $G$ is for $\gamma$ to index
a linear dependant set of rows of $J$. This is precisely
the cokernel of $J$.
\tombstone 

The result we are leading to characterizes robust classical codes 
as being equivalent to a row-reduction condition on the generator matrix.
The proof is by successive applications of the Pivot Lemma
and the Copivot Lemma.

\dotheorem{\numitem{th-robust}} (Robustness Theorem.)
Let $A$ be a classical code of length $n$, with generator matrix $G$.
Then $A$ is robust if and only if 
there is  some permutation of the columns of $G$ such that 
$G$ has a canonical form $G = (I_k\ \ J)$
with $J$ full rank.
\doproof 
Firstly, to prove the ``if'' part, we
assume that  $G = (I_k\ \ J)$ with $J$ full rank. 
We will find disjoint sets $\delta,\gamma\subset[n]$
that simultaneously puncture $G$ and $H$.

By Lemma~\ref{lemma:normal}, this code has parity check matrix
    $$ H = (J^\top \ \ I_m) $$
where $m=n-k.$
Choose $\gamma=[k]$, so that $\gamma$ is supported on the pivots of $G$ and
the copivots of $H$.
Because $J$ is full rank, the cokernel of $J$ is trivial and so 
by the Pivot Lemma we have that $\gamma$ punctures $G$.
From the Copivot Lemma we have that $\gamma$ punctures $H$.

Also by the Copivot Lemma we have any $\delta\subset\{k+1, ..., m\}$
punctures $G$, we just need to find such a $\delta$,
with $|\delta|=k$, that also punctures $H$.
This $\delta$ is a subset of the pivots of $H$ so we 
use the Pivot Lemma to see that $\delta$ must also puncture 
$\Cokernel(J^\top)=\Kernel(J)^\top.$
We show how to achieve this.
The matrix $J$ is full rank
so from Eq.~(\ref{eq:euler}) a matrix for 
$\Cokernel(J^\top)$ has $m-k$ rows and $m$ columns.
This $(m-k)\times m$ matrix is full rank and so 
has $m-k$ pivots and $k$ copivots.
We therefore choose $\delta$ to be these $k$ copivots of $\Cokernel(J^\top)$
and by the Copivot Lemma, $\delta$ will puncture $\Cokernel(J^\top)$.

Secondly,
to prove the ``only if'' part, assume we have 
disjoint sets $\gamma,\delta\subset[n]$
that simultaneously puncture $G$ and $H$.
The goal is to find a canonical form 
$G = (I_k\ \ J)$ with $J$ full rank. 
To do this, permute the columns of $G$ and $H$ so that 
$\gamma=\{1,...,k\}$ and $\delta=\{k+1,...,2k\}.$
With this column order, row-reduce $G$ to produce
the form $G = (I_k\ \ J)$.
Using the Pivot lemma, we see that $\gamma=\{1,...,k\}$ punctures
$\Cokernel(J)$ which implies this cokernel is trivial, and so $J$ is full-rank.
\tombstone 

%
%

\section{Discussion}

We have shown that this ensemble of good quantum codes nevertheless fails to
support transversal gates beyond the Clifford group.
This no-go result rules out a wide class of codes, and so assists
the search (construction) of codes that do have non-Clifford transversal gates.

The toric code, which is a hypergraph product code, does not satisfy 
the hypothesis of Theorem~\ref{th:main}.
This theorem applies to codes which are a generalization of the surface code.
It is a natural question to ask if this result can be strengthened to cover these
other hypergraph product codes. 
However, the argument as it stands cannot be strengthened: we have found examples where 
the region $\gamma\cap\delta$ (see Figure~\ref{fig:intersect}) is not correctable,
even though it satisfies the hypothesis of the Union Lemma.
But it may be that a more refined argument does work: numerics suggest that the
support of specific commutators
$pqp^{-1}q^{-1}$ of logical operators $p,q$ is always correctable.

The reader may wonder exactly how ``mild'' is the robustness condition.
We numerically searched through the 
Gallagher code ensembles used for building hypergraph product codes
in~\cite{Kovalev2018,Roffe2020}, and did not find any non-robust instance.
Specifically, we found evidence that any $[n,k,d]$ Gallagher code with
$d\ge 3$ and $k\le n/2$ is robust.

Another natural question to ask is the existence of transversal gates for
higher degree hypergraph (homological) product of classical codes 
$A\tensor B\tensor C$, etc.~\cite{Quintavalle2020}.
We expect these codes also have a theory of taut logical operators.
Given the intricacy of the calculations in the present work, it seems
that better techniques, and deeper insights, are needed before these results
can be pushed further.

\section*{Acknowledgements}

We thank Armanda Quintavalle, Joschka Roffe, Earl Campbell
and Michael Vasmer for useful discussions.
This research was a part of the
QCDA project (EP/R043647/1)
which has received funding from the QuantERA
ERA-NET Cofund in Quantum Technologies implemented within
the European Union's Horizon 2020 Programme.

\bibliography{refs}{}
\bibliographystyle{abbrv}

\end{document}